%% file: EGNC_paper_xyz.tex
\documentclass{CEAS_EuroGNC_v1.0}

\usepackage[utf8]{inputenc}

\usepackage{graphicx}
\usepackage{amsmath}
\usepackage[version=4]{mhchem}
\usepackage{siunitx}
\usepackage{array}
\usepackage{longtable}
\usepackage{siunitx}
\usepackage{mathtools}
\usepackage{amsmath}
\usepackage{amssymb}
\usepackage{caption}
\usepackage{floatrow}   
\usepackage{subcaption}
\usepackage{bm}
\DeclareMathOperator{\diag}{diag} 

\title{Robust line-of-sight pointing control on-board a stratospheric balloon-borne platform}

\ifpdf
\hypersetup{pdfinfo={
Title={Robust line-of-sight pointing control on-board a stratospheric balloon-borne platform},
Author={Ervan Kassarian},
Keywords={CEAS EuroGNC; Stratospheric balloons ; Robust control ; Line-of-sight pointing control ; Multibody dynamics}
}}
\fi

\begin{document}
\maketitle

\begin{authorList}{4cm}
\addAuthor{Ervan Kassarian}{ISAE-Supaero, 31400, Toulouse, France. \emailAddress{ervan.kassarian@isae-supaero.fr}}
\addAuthor{Francesco Sanfedino}{ISAE-Supaero, 31400, Toulouse, France. \emailAddress{francesco.sanfedino@isae-supaero.fr}}
\addAuthor{Daniel Alazard}{ISAE-Supaero, 31400, Toulouse, France. \emailAddress{daniel.alazard@isae-supaero.fr}}
\addAuthor{Johan Montel}{CNES, 31400, Toulouse, france. \emailAddress{johan.montel@cnes.fr}}
\addAuthor{Charles-Antoine Chevrier}{CNES, 31400, Toulouse, france. \emailAddress{charlesantoine.chevrier@cnes.fr}}
\end{authorList}

\begin{abstract}
This paper addresses the lack of a general methodology for the controller synthesis of an optical instrument on-board a stratospheric balloon-borne platform, such as a telescope or siderostat, to meet pointing requirements that are becoming more and more stringent in the context of astronomy missions.
Most often in the literature, a simple control structure is chosen, and the control gains are tuned empirically based on ground testings.
However, due to the large dimensions of the balloon and the flight chain, experimental set-ups only involve the pointing system and the platform, whereas flight experience shows that the pointing performance is essentially limited by the rejection of the natural pendulum-like oscillations of the fully deployed system.
This observation justifies the need for a model that predicts such flight conditions that cannot be replicated in laboratory, and for an adequate methodology addressing the line-of-sight controller design.
In particular, it is necessary to ensure robust stability and performance to the parametric uncertainties inherent to balloon-borne systems, such as complex balloon's properties or release of ballast throughout the flight, especially since experimental validation is limited.
In this paper, a dynamical model of the complete system is proposed, based on a multibody approach and accounting for parametric uncertainties with Linear Fractional Transformations. The comparison with flight data shows that the frequency content of the platform's motion is accurately predicted.
Then, the robust control of the line-of-sight is tackled as a $\mathcal H_{\infty}$ problem that allows to reach the performance objectives in terms of disturbance rejection, control bandwidth and actuators limitations.
\end{abstract}




\newpage

\input{1_introduction}

\input{2_modeling_chain}

\input{3_modeling_pointing}

\input{4_control}

\input{5_conclusion}

\section*{Acknowledgments}
This work was funded by ISAE-SUPAERO and CNES (the French space agency).

\bibliography{library}

\end{document}

%% file: 1_introduction.tex
\section{Introduction}

Reusable and flexible in their deployment and operability, while also cheaper, more ecological and with faster development times than satellites \cite[chap.1.3]{Yajima2009b}, stratospheric balloons have been carrying heavy scientific payloads into the near-space environment for decades.
In particular, optical instruments benefit from the low atmospheric absorption at such altitudes, making stratospheric balloons suited for astronomy missions. 
Such missions typically have sub-arcsecond pointing precision requirements \cite{Montel2019,Rhodes2012,Young2012,Diller2015,Mendillo2019,Hamden2020}, and current developments aim at further improving the precision with fast-steering mirrors \cite{Howe2017,Romualdez2020}.
To address more and more stringent precision objectives, it is necessary to develop accurate dynamical models and specific control strategies.

The dynamics of balloon-borne flight chains include the torsion of the flight train around the vertical axis, which can be modeled with the bi-filar pendulum model \cite{Fissel2013,Aubin2017}, and the pendulum-like oscillations in the vertical planes, modeled with Lagrangian mechanics \cite{Nigro1985a,Ward2003a} or multibody approaches \cite{Quadrelli2004a,Romualdez2018}.
A complete dynamical model was proposed in \cite{Kassarian2021} based on Lagrangian mechanics, taking into account the torsion and pendulum dynamics as well as their coupling.
This model was compared to flight data and produced satisfying predictions of the dominant natural modes of the system. 
However, balloon systems involve uncertain or varying parameters. Indeed, the gondola generally carries ballast that initially represents a significant portion of the gondola's mass and that is released throughout the flight for altitude control, the balloon's properties are difficult to predict, and more generally, experimental validation is rarely available since it requires to deploy the whole system in flight.
To address the modeling of uncertain multibody systems like stratospheric balloons, a general multibody framework was developed in \cite{Kassarian2021a}. The parametric uncertainties are represented with an unknown and bounded operator $\Delta$ based on the Linear Fractional Transformations (LFT), as it enables tools like $\mu$-analysis and $\mathcal H_{\infty}$ synthesis \cite{Zhou1996}.
To the best of the author's knowledge, a model of balloon systems accounting for parametric uncertainties does not yet exist. 
The first contribution of this paper is a complete control-oriented model of the system based on the multibody framework presented in \cite{Kassarian2021a} and accounting for parametric uncertainties with Linear Fractional Transformations.
An empirical model of the frequency content of the wind disturbance is proposed based on flight data, and allows to validate the nominal model of the flight chain dynamics. The pointing system is also modeled, including line-of-sight estimation, sensors and actuators dynamics, and frequency content of the measurement noise.

Once an accurate dynamical model is obtained, it is possible to address the pointing control of the optical instrument on-board the gondola, generally a telescope \cite{DeWeese2006,Rhodes2012,Diller2015,Varga2015,Stuchlik2015} or siderostat \cite{Benford2012,Hawat1996,Hamden2020}. 
Most often, simple control structures such as Proportional-Integral-Derivative (PID) gains are chosen \cite{Quine2002,Ward2003a,Benford2012,Stuchlik2015,Romualdez2016,Aboobaker2018} and the control gains are tuned empirically based on ground testings \cite{DeWeese2006,Nakano2010,Rhodes2012,Shariff2014,Diller2014,SHOJI2016,Jones-Wilson2017a}. There exists no general model-based methodology for controller synthesis, and, more critically, experimental ground-based set-ups are not representative of the dynamics of the fully deployed system in flight, whereas flight experience proves that the line-of-sight control is essentially limited by the rejection of the natural modes of the flight chain excited by wind disturbances \cite{Montel2019}.
However, modern control techniques based on non-smooth optimization of an $\mathcal H_{\infty}$ problem provide powerful tools to tackle robust control of complex uncertain systems \cite{Zhou1996,Apkarian2015,Apkarian2015a}. In particular, the uncertainties are taken into account during the control design, which allows to optimize the controller with regard to worst-case parametric configurations and to avoid time-consuming simulation-based approaches like Monte-Carlo campaigns.
The second contribution is a general methodology to address the robust line-of-sight pointing control of an optical instrument on-board a stratospheric balloon. Based on the LFT model presented in this paper, the relative pointing error \cite{Ott2013,ECSS2011a} caused by wind and noise disturbances is minimized in the sense of the worst-case $\mathcal H_{\infty}$-norm, while also ensuring frequency-domain constraints on the sensitivity function and actuators limitations.

Although the presented modeling and control methodology has extent to stratospheric balloons in general, the paper relies on the case study of the Faint Intergalactic-medium Redshifted Emission Balloon (FIREBall), a joint NASA/CNES experiment.
The modeling of the flight chain is described in Section~\ref{sec:modeling}. Then, Section~\ref{sec:pointing} addresses the modeling of the pointing system. Finally, the robust control of the line-of-sight is performed in Section~\ref{sec:control}.

%% file: 2_modeling_chain.tex
\section{Modeling of the flight chain}

\label{sec:modeling}

\subsection{The FIREBall flight chain}

\begin{figure}[!ht]
	\centering
	\includegraphics[width=.5\columnwidth]{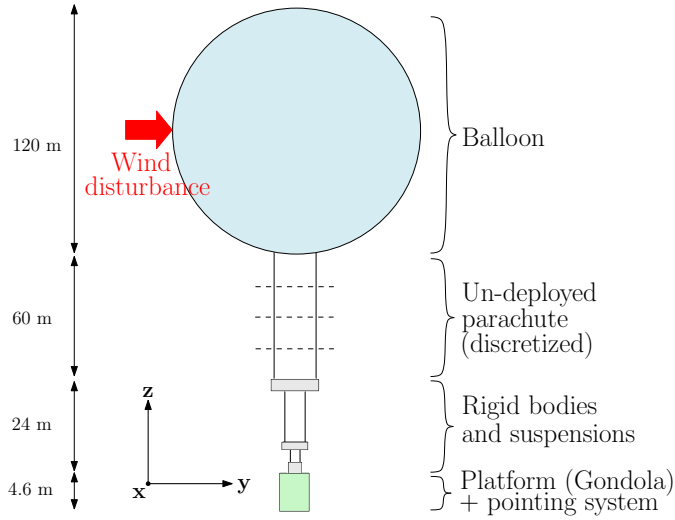}
	\caption{Sketch of the FIREBall system (not in scale)}
	 \label{fig:fireball}
\end{figure}

The stratospheric balloon FIREBall is represented in Fig.~\ref{fig:fireball}. The flight chain is decomposed in 11 elements, namely: the balloon, the parachute, which was discretized in 4 elements to account for its flexibility, various rigid bodies and filar suspensions, and the platform (gondola), carrying up to \SI{500}{\kilo\gram} of ballast and the pointing system which is detailed in Section \ref{sec:pointing}. The wind applies an aerodynamic force to the balloon that disturbs the system from its equilibrium state.

\subsection{Nominal model with Lagrangian mechanics}

\label{sec:nominal}

A complete dynamical model of balloon-borne flight chains was proposed in \cite{Kassarian2021}.
Generally speaking, the flight chain is considered to be composed of two types of bodies: rigid bodies (such as the platform) which can rotate in the three directions, and bi-filar suspensions, which are modeled as a single body with an additional degree of freedom corresponding to their torsion around $\mathbf z$. Note that the parachute is discretized and modeled as 4 bi-filar suspensions, and that the balloon is considered as a rigid body. The mechanisms between two successive bodies are revolute joints that allow the rotation around $\mathbf x$ and $\mathbf y$.
The modeling distinguishes two types of dynamics: (i) the pendulum-like dynamics, corresponding to the oscillations of the elements of the flight chain around $\mathbf x$ and $\mathbf y$, and (ii) the torsion dynamics, corresponding to the torsion of the bi-filar suspensions around the vertical axis $\mathbf z$.
With the small angles approximation, the pendulum and torsion dynamics are decoupled in open-loop, and the equations of motion are derived with Lagrangian mechanics:
\begin{equation} \label{eq:nominal_model}
\bm \mu \delta \ddot{\mathbf X} + \bm \gamma \delta \dot{\mathbf X} + \bm \kappa \delta \mathbf X = \delta \mathbf T
\end{equation}
where $\delta \mathbf X$ contains the degrees of freedom, $\delta \mathbf T$ is the vector of input forces and torques, the mass matrix $\bm \mu$ and the stiffness matrix $\bm \kappa$ are obtained following the method presented in \cite{Kassarian2021}, and the damping matrix $\bm \gamma$ assumes natural dissipative effects in the rotation of the balloon and in the revolute joints. Since the values of the damping coefficients are not theoretically known, they were empirically estimated from flight data (see Section \ref{sec:validation}).

\subsection{Disturbance model and validation based on flight data}

\label{sec:validation}

\begin{figure}[!ht]
	\centering
	\includegraphics[width=.9\columnwidth]{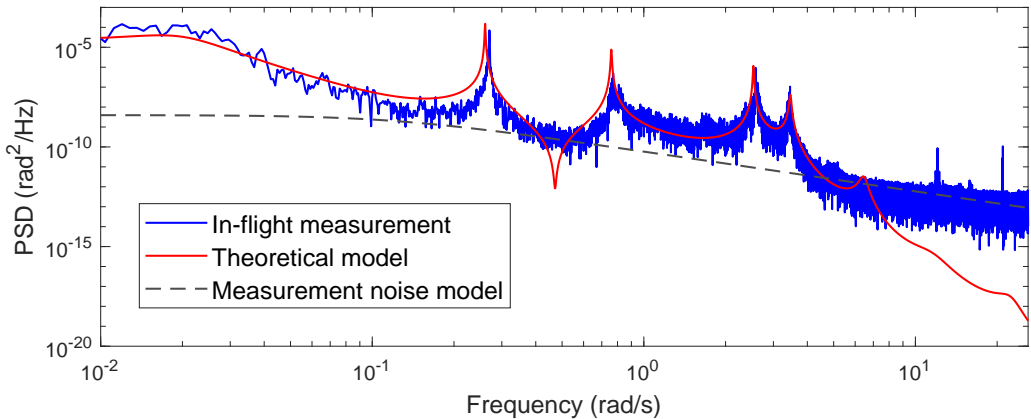}
	\caption{Platform's motion frequency content: model VS flight data}
	 \label{fig:modelvsdata}
\end{figure}

The wind disturbance is modeled as a colored noise.
Let us note $\delta F_y$ the force applied to the balloon's center of pressure along $\mathbf y$, and $D(\mathrm s)$ the filter representing its frequency content:
\begin{equation}
    \delta F_y (\mathrm s) =  D(\mathrm s) \delta d_y(\mathrm s)
\end{equation}
where $\delta d_y$ is a white noise input of unitary power spectral density (PSD): $\Phi_{\delta d_y} (j \omega)=1$ where the function $\Phi_x$ designates the PSD of a signal $x$.
From the nominal model presented in equation \eqref{eq:nominal_model}, let us note $G(\mathrm s)$ the transfer function from $\delta F_y$ to the platform's angle $\theta^p_x$ about $\mathbf x$. Assuming that the aerodynamic force is the sole contributor to the platform's oscillations, the PSD of $\theta^p_x$ reads:
\begin{equation}
    \Phi_{\theta^p_x} (j \omega) = \mid D(j \omega) G(j\omega) \mid ^2 \;.
\end{equation}
The function $D(\mathrm s)$ was empirically determined to fit the general shape of the PSD $\Phi_{\theta^p_x} (j \omega)$ computed from flight data:
\begin{equation} \label{eq:disturbance_model}
    D(\mathrm s) = 4.62 \frac{\mathrm s^2 + 0.21 \mathrm s + 0.0225}{\mathrm s^2 + 0.016 \mathrm s + \SI{4e-4}{}} \;.
\end{equation}
Then, the damping coefficients in the matrix $\bm \gamma$ (see Section \ref{sec:nominal}) were adjusted to fit the amplitude of the resonances observed at \SI{0.27}{\radian\per\second}, \SI{0.76}{\radian\per\second}, \SI{2.5}{\radian\per\second} and \SI{3.5}{\radian\per\second} respectively, corresponding to the first four pendulum modes of the system. The result is shown in Fig.~\ref{fig:modelvsdata}, with the PSD $\Phi_{\theta^p_x} (j \omega)$ computed from in-flight measurement (in blue), the theoretical model $\mid D(j \omega) G(j\omega) \mid ^2$ (in red), and the PSD of the measurement noise (in grey dotted line). The model fits up to \SI{5}{\radian\per\second}. For higher frequencies, the signal is dominated by the measurement noise and no conclusion can be made.
The same filter $D(\mathrm s)$ is used to model the aerodynamic force $\delta F_x$ along $\mathbf x$. The fit with $\Phi_{\theta^p_y} (j \omega)$ is very similar to Fig.~\ref{fig:modelvsdata}, and is not presented here.

Remark: the model $D(\mathrm s)$ obtained in equation \eqref{eq:disturbance_model} yields disturbances of higher amplitude than Dryden wind gusts models (with a factor $\sim$10), which are widely used in aircraft design \cite{Hoblit1988,Defense2004}. It is suspected that the aerodynamic disturbances encountered during the balloon flight originate from the variations of the wind correlated with altitude oscillations \cite{Treilhou2000,Alexander2011}. This is also suggested by the cut-off frequency of $D(\mathrm s)$ that coincides with the frequency of the buoyant oscillations. However, the model presented in this paper is purely empirical based on data of the FIREBall 2018 flight. A general disturbance model would most likely depends on the altitude of the mission and on the balloon properties.

\subsection{Uncertain LFT model with a multibody approach}

\label{sec:lftmodel}

The nominal model in Section \ref{sec:nominal} alongside with the disturbance model in Section \ref{sec:validation} yield a good prediction of the gondola's motion in terms of amplitude and frequency content in nominal conditions. However, it is necessary to account for parametric uncertainties which may modify the system's behavior. Indeed: (i) the balloon's characteristics may vary throughout the flight, (ii) the ballast mass is initially \SI{500}{\kilo\gram}, but decreases down to 0 during the flight, (iii) various parametric uncertainties may explain the small discrepancies between the model and the flight data in Fig.~\ref{fig:modelvsdata}, and (iv) flight data is not always available to validate the model.

Generating a Linear fractional Transformation (LFT) model from the nominal model in equation~\eqref{eq:nominal_model} is not a viable option since it yields many redundant occurrences of the uncertain parameters, which hinders the robust controller synthesis. Therefore, a general approach was developed in \cite{Kassarian2021a} to compute LFT models of multibody systems.
In this approach, each substructure (body, kinematic joint...) has its individual LFT model based on Newton-Euler equations, expressing the relationship between the motion of the substructure and the wrench (force and torque) applied to it. These individual models are then assembled to build the complex multibody structure while taking care of the dependency of the trim conditions to the uncertain parameters.  The LFT model of the structure is finally obtained as a continuous function of the uncertain parameters, with a limited number of occurrences, and covers all plants within the specified bounds in a single model, without introducing conservatism.

\begin{figure}[!ht]
\centering
\begin{subfigure}{.4\linewidth}
    \centering
    \includegraphics[width=.8\linewidth]{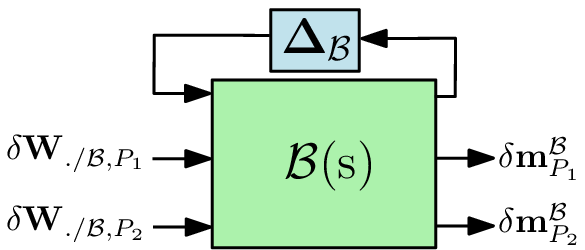}
    \caption{Rigid body (forward dynamics)}\label{fig:indiv1}
\end{subfigure}
\begin{subfigure}{.4\linewidth}
    \centering
    \includegraphics[width=.8\linewidth]{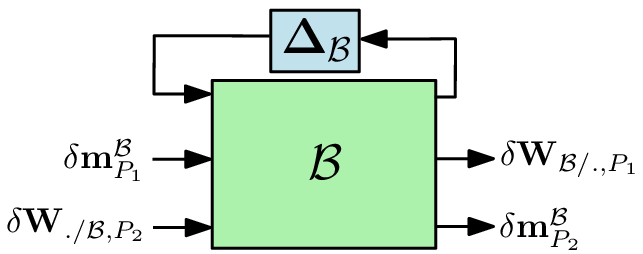}
    \caption{Rigid body (inverse dynamics)}\label{fig:indiv2}
\end{subfigure}

\bigskip
\begin{subfigure}{.4\linewidth}
    \centering
    \includegraphics[width=.8\linewidth]{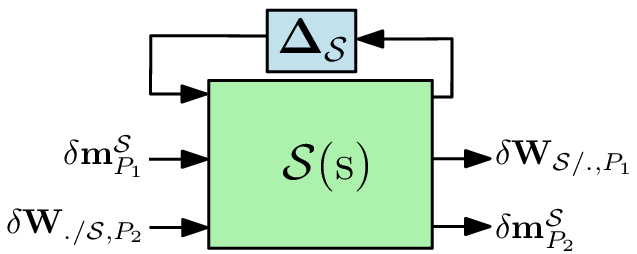}
    \caption{Bi-filar suspension}\label{fig:indiv3}
\end{subfigure}
\begin{subfigure}{.4\linewidth}
    \centering
    \includegraphics[width=.8\linewidth]{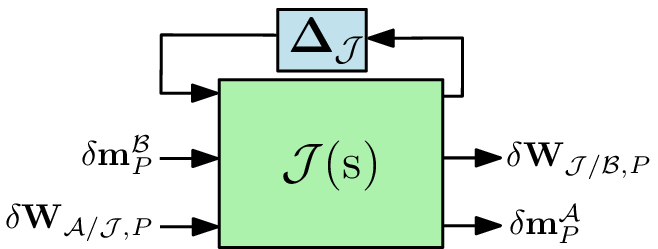}
    \caption{Revolute joint}\label{fig:indiv4}
\end{subfigure}
\caption{Individual models constituting the balloon system \label{fig:individual_models}}
\end{figure}

This approach was applied to the FIREBall system. 
Considering the linear dynamics around the equilibrium, let us call ``wrench'' the vector $\delta \mathbf W_{\mathcal A/\mathcal B,P} \in \mathbb R^{6}$ containing the force and torque applied by a body $\mathcal A$ to a body $\mathcal B$ at point $P$, and ``motion vector'' the vector $\delta \mathbf m^{\mathcal B}_{P} \in \mathbb R^{18}$ containing the linear and angular accelerations, speeds and positions of body $\mathcal B$ at point $P$. 
The individual linearized models of the bodies (rigid bodies, bi-filar suspensions and revolute joints) constituting the system are represented in Fig.~\ref{fig:individual_models}:
\begin{itemize}
    \item In the forward dynamics model of a rigid body (Fig.~\ref{fig:indiv1}), the inputs are the wrenches $\delta \mathbf W_{./\mathcal B,P_1}$ and $\delta \mathbf W_{./\mathcal B,P_2}$ applied to $\mathcal B$ at connection points $P_1$ and $P_2$. The model returns the motion vectors $\delta \mathbf m^{\mathcal B}_{P_1}$ and $\delta \mathbf m^{\mathcal B}_{P_2}$ at the connection points $P_1$ and $P_2$. The forward dynamics model is a 12-th order dynamical model (the wrenches are integrated twice to produce the motion vector).
    \item In the inverse dynamics model of a rigid body (Fig.~\ref{fig:indiv2}), the motion vector $\delta \mathbf m^{\mathcal B}_{P_1}$ is imposed as input at one connection point $P_1$, and a wrench $\delta \mathbf W_{./\mathcal B,P_2}$ is applied at point $P_2$. The model returns the wrench $\delta \mathbf W_{\mathcal B/.,P_1}$ applied by the body in reaction at $P_1$ and the motion vector $\delta \mathbf m^{\mathcal B}_{P_2}$ at $P_2$. The inverse dynamics model is a static model.
    \item The model of the bi-filar suspension (Fig.~\ref{fig:indiv3}) has the same inputs and outputs as the inverse dynamics model of a rigid body. However, the bi-filar suspension model has an additional degree of freedom corresponding to the twist around the vertical axis (see \cite{Kassarian2021} for the bi-filar suspension model based on Lagrangian mechanics). It is a 2-nd order dynamic model.
    \item The revolute joint connects two given bodies noted $\mathcal A$ and $\mathcal B$ at connection point $P$ with a degree of freedom in rotation. It is a 2-nd order dynamic model.
\end{itemize}
Each nominal model (green blocks) is augmented with parametric uncertainties, represented by the unknown, bounded operators $\bm \Delta_{(.)}$ (blue blocks). In particular, the blocks $\bm \Delta_{\mathcal J}$ in the revolute joints correspond to uncertainties in the stiffness due to uncertain masses in the flight chain. The reader is referred to \cite{Kassarian2021a} for more details on the derivation of these individual linearized models and their assembly procedure with parameter-dependent trim conditions.

\begin{figure}[!ht]
	\centering
	\includegraphics[width=.28\columnwidth]{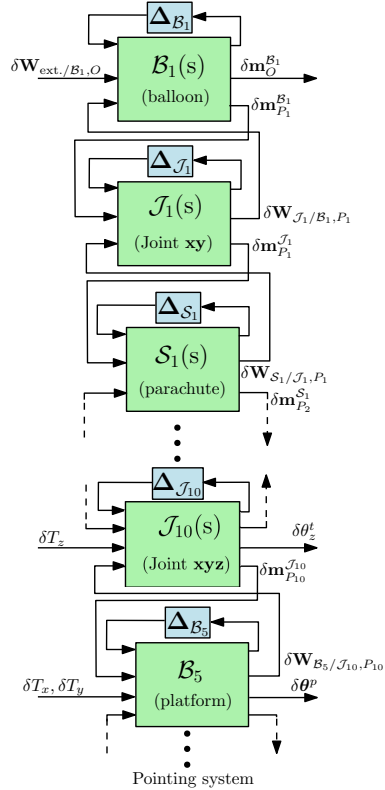}
	\caption{Multibody LFT model of the flight chain}
	 \label{fig:multibody_balloon}
\end{figure}

Finally, the linearized model of the flight chain composed of 5 rigid bodies (the balloon is modeled with a forward dynamics model, and all other rigid bodies with an inverse dynamics model), 6 bi-filar suspensions, and 10 joints (at connection points $P_1$ to $P_{10}$), is represented in Fig.~\ref{fig:multibody_balloon}. The joints $\mathcal J_{1}$ to $\mathcal J_{9}$ allow the rotation around $\mathbf x$ and $\mathbf y$. $\mathcal J_{10}$ can additionally be actuated by a torque $\delta T_z$ around the vertical axis $\mathbf z$ to model the azimuth control, and the torques $\delta T_x$ and $\delta T_y$ can be applied to the platform using reaction wheels (see section \ref{sec:primary_control}). The output vector $\delta \bm \theta^p$ contains the angles of the platform, and the output scalar $\delta \theta^t_z$ is the flight train's angle around $\mathbf z$. The input $\delta \mathbf W_{\mathrm{ext}./\mathcal B_1,O}$ contains the aerodynamic disturbances (input), the damping terms (added as feedback on the rotation rates of the balloon), and the variations of the buoyancy force projected in the reference frame attached to the balloon (added as feedback on the angular position of the balloon).
The parametric uncertainties concern: (i) the gondola's parameters: moments of inertia ($\pm5\%$), mass ($\pm5\%$), center of gravity ($\pm0.1$m), ballast mass (comprised between 0 and \SI{500}{\kilo\gram}); (ii) the balloon's parameters: moments of inertia ($\pm5\%$), mass ($\pm5\%$), center of pressure ($\pm3$m), center of gravity ($\pm3$m), center of buyoancy ($\pm3$m) ; and (iii) other more minor parameters of the flight chain: radius of the parachute ($\pm10\%$), mass of the body $\mathcal B_2$ ($\pm10\%$). 
It is verified that the nominal value of this uncertain model exactly matches the nominal model obtained in Section \ref{sec:nominal}.

\subsection{Primary control of the platform}

\label{sec:primary_control}

The primary control ensures that the azimuth of the platform (angle around the vertical axis $\mathbf z$) follows the sideral motion of the target. Since the associated pointing requirements are not very demanding, the controller synthesis of the primary control is not detailed in this paper. However, it is necessary to take it into account in the flight chain model before addressing the line-of-sight pointing control.

The primary control in azimuth is performed with a motorized pivot, which applies a torque $\delta T_z$ to the gondola and the opposite torque $-\delta T_z$ to the bottom of the flight chain. The torque is computed with two 2-nd order controllers $C_1(\mathrm s)$ and $C_2(\mathrm s)$ using the reference azimuth $\delta r_{az}$, the platform's angle $\delta \theta^p_z$ and the flight train's angle $\delta \theta^t_z$ so as to provide a \SI{1}{\radian\per\second} bandwidth while damping the torsion modes of the flight train:
\begin{equation}
     \delta T_z = C_1(\mathrm s) (\delta r_{az} - \delta \hat{\theta}^p_z) + C_2(\mathrm s) \delta \hat{\theta}^t_z
\end{equation}
with the estimated angles
\begin{equation}
\left\lbrace
\begin{array}{l}
     \delta \hat{\theta}^p_z = d(\mathrm s) \delta \theta^p_z \\
     \delta \hat{\theta}^t_z = d(\mathrm s) \left(\frac{40 \mathrm s}{40 \mathrm s + 1}\right)^2 \delta \theta^t_z
\end{array} \right.
\end{equation}
where $d(\mathrm s)$ designates a third-order Pad\'e approximation of a \SI{70}{\milli\second} delay, a low-pass filter is applied to the flight train's angle $\delta \theta^t_z$ (to avoid low-frequency drift, and because only the platform should be controlled in low-frequency), and the controllers
\begin{equation}
        \left\lbrace \begin{array}{l}
         C_1(\mathrm s) = \frac{3043 \mathrm s^2 + 958000 \mathrm s + 7360}{\mathrm s^2 + 357 \mathrm s + 900} \\
         C_2(\mathrm s) = \frac{140 \mathrm s^2 + 3259 \mathrm s + 166}{\mathrm s^2 + 357 \mathrm s + 900}
    \end{array} \right. \;.
\end{equation}

In addition to the azimuth control, the FIREBall system carries two reaction wheels able to generate a torque about $\mathbf x$ and $\mathbf y$ respectively. Taking the rotation rates of the platform $\delta \dot \theta^p_x$ and $\delta \dot \theta^p_y$ as inputs, the controller $C_3(\mathrm s)$ is tuned to provide active damping to the pendulum modes 3 and 4, because they play an important role in the pointing performance \cite{Montel2019}. The torques generated by the wheel read:
\begin{equation}
    \left[ \begin{array}{c}
         \delta T_x \\
         \delta T_y
    \end{array} \right]
     = - \underbrace{\frac{0.885 \mathrm s}{\mathrm s + 1.847}}_{\text{Actuator model}} \underbrace{\frac{1233 \mathrm s + 963}{\mathrm s^2 + 0.76 \mathrm s + 6.3}}_{C_3(\mathrm s)} d(\mathrm s)
         \left[ \begin{array}{c}
         \delta \dot \theta^p_x \\
         \delta \dot \theta^p_y
    \end{array} \right] \;.
\end{equation}

%% file: 3_modeling_pointing.tex
\section{Modeling of the pointing system}

\label{sec:pointing}

\subsection{Description of the pointing system}

\label{sec:description_system}

\begin{figure}[!ht]
	\centering
	\includegraphics[width=.4\columnwidth]{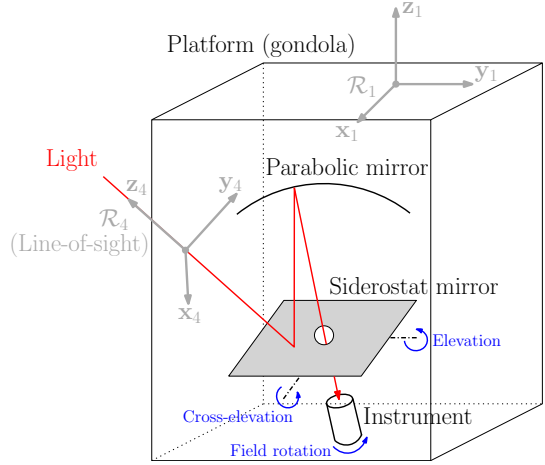}
	\caption{Sketch of the pointing system}
	 \label{fig:pointing_system}
\end{figure}

The purpose of the pointing system represented in Fig.~\ref{fig:pointing_system} is to reject the gondola's motion (up to \SI{0.15}{\degree} of amplitude \cite{Montel2019}) from the line-of-sight to obtain a pointing accuracy around the arc second (elevation and cross-elevation axes) and arc minute (field rotation axis).
The siderostat mirror is mounted on a gimbal equipped with two DC motors which allow to control it around two axes called elevation and cross-elevation.
The light is reflected on the siderostat mirror, and then concentrated by a parabolic mirror to the instrument through a hole in the siderostat mirror. The instrument is mounted on a rotating stage, allowing to control a third axis called field rotation.
Let us note $\mathcal R_1=(\mathbf x_1, \mathbf y_1,\mathbf z_1)$ the reference frame attached to the platform (gondola).
The line-of-sight, represented with the frame $\mathcal R_4=(\mathbf x_4, \mathbf y_4,\mathbf z_4)$, is defined with the three successive rotations:
\begin{itemize}
    \item A first rotation around $\mathbf y_1$ defines the frame $\mathcal R_2=(\mathbf x_2, \mathbf y_2,\mathbf z_2)$. The angle between $\mathcal R_1$ and $\mathcal R_2$ is called elevation and equals $\SI{10}{\degree} + 2\theta_{\mathrm{el}}$, where $\theta_{\mathrm{el}}$ is the angle of the siderostat mirror around $\mathbf y_1$. The factor 2 is due to the reflection of the light on the siderostat mirror, and the term \SI{10}{\degree} is due to the angular position of the parabolic mirror.
    \item A second rotation around $\mathbf x_2$ defines the frame $\mathcal R_3=(\mathbf x_3, \mathbf y_3,\mathbf z_3)$. The angle between $\mathcal R_2$ and $\mathcal R_3$ is called cross-elevation and equals $2\theta_{\mathrm{ce}}$, where $\theta_{\mathrm{ce}}$ is the angle of the siderostat mirror around $\mathbf x_2$. The factor 2 is due to the reflection of the light on the siderostat mirror.
    \item A third rotation around $\mathbf z_3$ defines the frame $\mathcal R_4=(\mathbf x_4, \mathbf y_4,\mathbf z_4)$ attached to the line-of-sight. The angle between $\mathcal R_3$ and $\mathcal R_4$ is called field rotation and is noted $\theta_{\mathrm{fr}}$.
\end{itemize}

\subsection{Definition of the line-of-sight}

\label{sec:los}

Let us note $\bm \theta^{EUL}$ the vector containing the Euler angles (in any given sequence) defining the orientation of $\mathcal R_4$.  It can be expressed as a function of the vector $\bm \theta^{p}$ containing the Euler angles of the platform, and of the vector $\bm \theta^{a} = \begin{bsmallmatrix} \theta_{\mathrm{el}} & \theta_{\mathrm{ce}} & \theta_{\mathrm{fr}} \end{bsmallmatrix}^T$ containing the angles of the three rotations imposed by the actuators (elevation, cross-elevation, field rotation) to the mirror or instrument. 
To the first order, this relationship reads:
\begin{equation}
    \delta \bm \theta^{EUL} = \left. \frac{\partial \bm \theta^{EUL}}{\partial \bm \theta^{p}} \right|_{\mathrm{eq}} \delta \bm \theta^{p} + \left. \frac{\partial \bm \theta^{EUL}}{\partial \bm \theta^{a}}\right|_{\mathrm{eq}} \mathbf R \delta \bm \theta^{a}
\end{equation}
where the matrix $\mathbf R = \diag(2,2,1)$ accounts for the reflections on the elevation and cross-elevation axes, and the $3 \times 3$ matrices $\frac{\partial \bm \theta^{EUL}}{\partial \bm \theta^{p}}$ and $\frac{\partial \bm \theta^{EUL}}{\partial \bm \theta^{a}}$ account for the rotations and depend on the equilibrium configuration. 
At equilibrium, the actuator's elevation angle $\bar \theta_{\mathrm{el}}$ is comprised between \SI{5}{\degree} and \SI{20}{\degree}, corresponding to a LOS elevation angle comprised between \SI{20}{\degree} and \SI{50}{\degree} (as defined in Section \ref{sec:description_system}), and the cross-elevation and field rotation angles are zero.
It is important to account for the different values that the equilibrium elevation angle $\bar \theta_{\mathrm{el}}$ can have during the flight, since it changes the influence of the platform's motion, and consequently of the wind disturbance, on the cross-elevation and field rotation axes. Therefore, $\bar \theta_{\mathrm{el}}$ is considered as an uncertain parameter in this study. As discussed in \cite{Dubanchet2016,Kassarian2021a}, it is necessary to define the uncertain parameter $t_{\mathrm{el}} = \mathrm{tan}\left( \bar \theta_{\mathrm{el}}/2 \right)$, instead of $\bar \theta_{\mathrm{el}}$, because the sine and cosine functions can be expressed as rational expressions in the variable $t_{\mathrm{el}}$.

Provided that $\frac{\partial \bm \theta^{EUL}}{\partial \bm \theta^{a}}$ is invertible, let us define the new vector $\delta \bm \theta^{LOS}$ describing the motion of the line-of-sight in elevation, cross-elevation and field rotation, such that:
\begin{equation} \label{eq:angles_los}
    \delta \bm \theta^{LOS} = \left. \frac{\partial \bm \theta^{EUL}}{\partial \bm \theta^{a}}\right|_{\mathrm{eq}}^{-1} \delta \bm \theta^{EUL} = \underbrace{\left. \frac{\partial \bm \theta^{EUL}}{\partial \bm \theta^{a}}\right|_{\mathrm{eq}}^{-1} \left.\frac{\partial \bm \theta^{EUL}}{\partial \bm \theta^{p}}\right|_{\mathrm{eq}}}_{=\mathbf P(t_{\mathrm{el}}) = \mathcal F_u (\mathbf P, \bm \Delta_{t_{\mathrm{el}}})} \delta \bm \theta^{p} + \mathbf R \delta \bm \theta^{a} \;.
\end{equation}
The vector $\delta \bm \theta^{LOS}$ is introduced because (i) the science requirements and the levels of noise of the optical sensor are expressed with this convention, and (ii) the action of the actuators on the line-of-sight angles are decoupled in equation \eqref{eq:angles_los} which facilitates the controller design.
The matrix $\mathbf P(t_{\mathrm{el}}) = \mathcal F_u (\mathbf P, \bm \Delta_{t_{\mathrm{el}}})$ expresses the relation between the gondola's motion and the LOS depending on $t_{\mathrm{el}}$, where $\mathcal F_u(.)$ designates the upper LFT operator.

\subsection{Estimation of the line-of-sight}

Three sets of sensors are used to estimate the line-of-sight:
\begin{itemize}
    \item An optical sensor is attached to the instrument and provides a direct measurement of the line-of-sight, noted $\delta \widehat{\bm \theta}^{LOS}_{\mathrm{optic}}$. It has low noise (around $\pm$0.1 arcsec in elevation and cross-elevation, $\pm$10 arcsec in field rotation, at \SI{30}{Hz}) but has a delay of \SI{40}{\milli\second}, due to the exposure time of the sensor and some processing delays, which limits the closed-loop performance and stability.
    \item Another estimate, noted $\delta \widehat{\bm \theta}^{LOS}_{\mathrm{gyro}}$, is calculated with equation \eqref{eq:angles_los} from the two measurements:
    \begin{itemize}
        \item Analogical gyrometers deliver the rotation rates of the platform with high noise (around $\pm$20 arcsec/s at \SI{200}{\hertz}) but no delay. These rotation rates are integrated to get an estimate $\delta \widehat{\bm \theta}^{p}$ of the attitude of the gondola.
        \item The position of each actuator is measured relatively to the gondola with negligible noise (with regard to the attitude provided by the gyrometers) and a 1 ms delay. This measurement is noted $\delta \widehat{\bm \theta}^{a}$.
    \end{itemize}
\end{itemize}

The line-of-sight measured by the optical sensor $\delta \widehat{\bm \theta}^{LOS}_{\mathrm{optic}}$ provides a good estimate at low frequencies, but cannot be used in high frequencies due to the delay. On the contrary, the gyrometers/actuators estimation $\delta \widehat{\bm \theta}^{LOS}_{\mathrm{gyro}}$ cannot be used in low frequencies due to the integration of the rate measurement resulting in high noise levels. Therefore, the optical sensor and the gyrometers/actuators measurements are filtered to provide an estimate $\delta \widehat{\bm \theta}^{LOS}$:

\begin{equation}
    \delta \widehat{\bm \theta}^{LOS} = \left( \mathbf I_3 - \mathbf F(\tau_1, \tau_2, \mathrm s) \right) \delta \widehat{\bm \theta}^{LOS}_{\mathrm{optic}} +  \mathbf F(\tau_1, \tau_2, \mathrm s) \delta \widehat{\bm \theta}^{LOS}_{\mathrm{gyro}}
\end{equation}
where the filter $\mathbf F(\tau_1, \tau_2, \mathrm s)$ is expressed as a LFT ($\mathcal F_l(.)$ designates the lower LFT operator):
\begin{equation} \label{eq:filter}
    \mathbf F(\tau_1, \tau_2, \mathrm s) = \frac{\tau_1 \mathrm s}{\tau_1 \mathrm s + 1} \frac{\tau_2 \mathrm s}{\tau_2 \mathrm s + 1} \mathbf I_3 = \mathcal F_l \left( \mathbf F(\mathrm s), \begin{bsmallmatrix} \tau_1 \mathbf I_3 & \mathbf 0  \\ \mathbf 0 & \tau_2 \mathbf I_3
\end{bsmallmatrix} \right) \;.
\end{equation}
The structure of the filter presented in equation \eqref{eq:filter} was fixed for the mission, but the two parameters $\tau_1$ and $\tau_2$ can be tuned to address the trade-off between:
\begin{itemize}
    \item noise levels: small values of $\tau_1$ and $\tau_2$ yield lower noise -- because the estimate of the LOS relies more on the optical measurement (small noise, but large delay),
    \item stability and rejection of the wind disturbance: high values of $\tau_1$ and $\tau_2$ yield good stability and disturbance rejection properties -- because the estimate of the LOS relies more on the gyrometer/actuators measurement (small delay, but high noise).
\end{itemize}
This trade-off is coupled with the controller synthesis. Therefore, the parameters $\tau_1$ and $\tau_2$ will be tuned simultaneously with the controller in Section \ref{sec:control}.

\begin{figure}[!ht]
	\centering
	\includegraphics[width=.6\columnwidth]{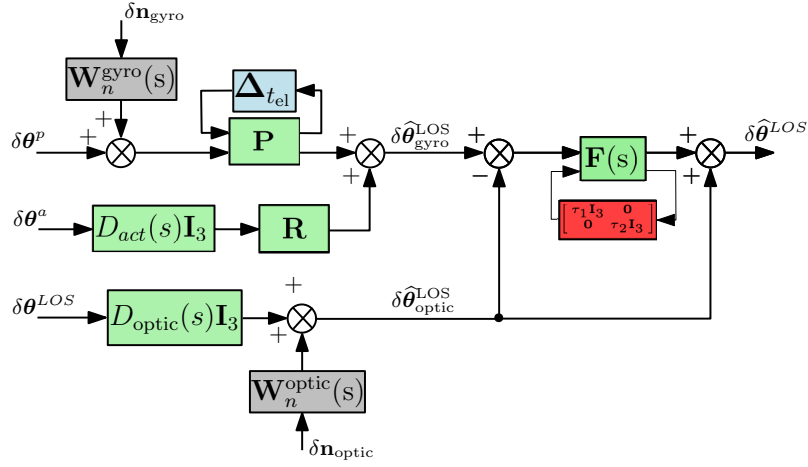}
	\caption{Line-of-sight estimation }
	 \label{fig:estimation}
\end{figure}

The block diagram in Fig.~\ref{fig:estimation} represents the estimation described in this section and implemented for controller synthesis. The transfer functions of the delays are treated as 3-rd order Padé approximations -- noted $D_{\mathrm{act}}(\mathrm s)$ for the actuators' sensors and $D_{\mathrm{optic}}(\mathrm s)$ for the optical measurement. The measurement noise of the optical sensor is treated as a white noise in the frequency range \SI{0}{\hertz} -- \SI{30}{\hertz} ; the measurement noise of the gyrometer is treated as a white noise in the frequency range \SI{0}{\hertz} -- \SI{200}{\hertz}, which is integrated to provide the position. The filters modeling the noise frequency contents are:

\begin{equation} \label{eq:noise}
    \left\lbrace \begin{array}{l}
          \mathbf W_n^{\mathrm{gyro}} = \SI{5e-6}{} \frac{\mathbf I_3}{\mathrm s} \\
          \mathbf W_n^{\mathrm{optic}} = \SI{4.5e-8}{}  \diag \left(1,1,100    \right)
    \end{array} \right. \;.
\end{equation}

\subsection{Actuation system}

Each actuator is represented by a transfer function $\delta \ddot \theta/\delta V$ between the input voltage and the angular acceleration imposed to the mirror or instrument. 
The transfer functions are given in table \ref{tab:actuators}, and the parameters $\tau_{\mathrm{el}}$, $\tau_{\mathrm{ce}}$ and $\tau_{\mathrm{fr}}$ each have 10\% uncertainty around their respective nominal values \SI{32}{\milli\second}, \SI{16}{\milli\second} and \SI{50}{\milli\second} respectively.

\begin{table}[hbt!]
\caption{\label{tab:actuators} Transfer functions of the actuators}
\centering
\resizebox{\columnwidth}{!}{%
\begin{tabular}{c|c|c|c}
 & Elevation & Cross-elevation & Field rotation \\ \hline
$\delta \ddot \theta/\delta V$ & $18800 \frac{\mathrm s + 10^{-3}}{\tau_{\mathrm{el}} \mathrm s + 1} = \mathcal F (A_{\mathrm{el}} (\mathrm s), \Delta_{\tau_{\mathrm{el}}})$ & $4200 \frac{\mathrm s + 10^{-3}}{\tau_{\mathrm{ce}} \mathrm s + 1} = \mathcal F (A_{\mathrm{ce}} (\mathrm s), \Delta_{\tau_{\mathrm{ce}}})$ & $1000 \frac{\mathrm s + 10^{-3}}{\tau_{\mathrm{fr}} \mathrm s + 1} = \mathcal F (A_{\mathrm{fr}} (\mathrm s), \Delta_{\tau_{\mathrm{fr}}})$ 
\end{tabular}
}
\end{table}

\begin{figure}[!ht]
	\centering
	\includegraphics[width=.3\columnwidth]{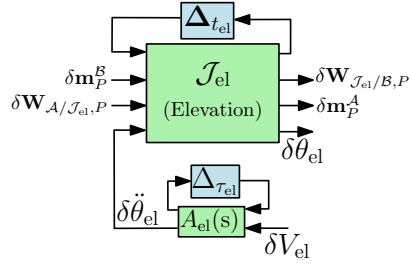}
	\caption{Elevation actuation system}
	 \label{fig:actuator_el}
\end{figure}

Furthermore, a revolute joint is used to model the rotation of the mirror or instrument.
Contrary to the revolute joints used in the modeling of the flight chain (see Fig.~\ref{fig:indiv4} in Section \ref{sec:lftmodel}), the revolute joints used in the pointing system are actuated. Therefore, they have an additional input, that is the angular acceleration $\delta \ddot \theta$ imposed by the actuator between the two bodies designated with the generic notations $\mathcal A$ and $\mathcal B$. The angular configuration $\delta \theta$ is returned as output.
For example, the elevation actuation system is represented in Fig.~\ref{fig:actuator_el}. The input voltage is noted $\delta V_{\mathrm{el}}$ and the angular configuration is noted $\delta \theta_{\mathrm{el}}$. The nominal model of the joint is noted $\mathcal J_{\mathrm{el}}$, and the block $\bm \Delta_{t_{\mathrm{el}}}$ represents the uncertainties due to the elevation angle at equilibrium $\bar \theta_{\mathrm{el}}$.

\begin{figure}[!ht]
	\centering
	\includegraphics[width=.7\columnwidth]{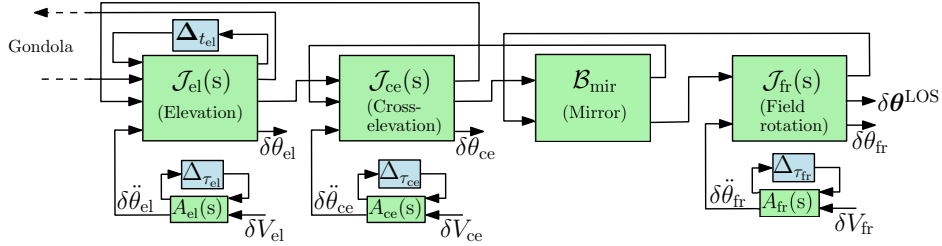}
	\caption{3-axes pointing system}
	 \label{fig:actuators}
\end{figure}

Finally, the pointing system is represented in Fig.~\ref{fig:actuators}.
The mirror is a rigid body with a matrix of inertia $\mathbf J = \diag(30, 30, 50) \; \mathbf I_3$ \SI{}{\kilo\gram\square\meter}. Note that its mass was already taken into account in the mechanical model of the gondola, since it influences its total mass, moments of inertia and position of its center of gravity. The variations of these quantities with the angular position of the mirror are within the uncertainty bounds set in Section \ref{sec:lftmodel}.
The inertia of the instrument in field rotation is negligible, and thus it is not represented.
The interconnection signals are not named for readability -- see Fig.~\ref{fig:actuator_el} for the details on a single axis actuation system.
The input voltages are noted $\delta V_{\mathrm{el}}$, $\delta V_{\mathrm{ce}}$ and $\delta V_{\mathrm{fr}}$ for the elevation, cross-elevation and field rotation actuators respectively. The outputs of the system are the angle of each actuator, respectively $\delta \theta_{\mathrm{el}}$, $\delta \theta_{\mathrm{ce}}$ and $\delta \theta_{\mathrm{fr}}$, and the line-of-sight angles $\delta \bm \theta^{\mathrm{LOS}}$.

%% file: 4_control.tex
\section{Robust pointing control of the line-of-sight}

\label{sec:control}

\subsection{Problem formulation}

\begin{figure}[!ht]
	\centering
	\includegraphics[width=.5\columnwidth]{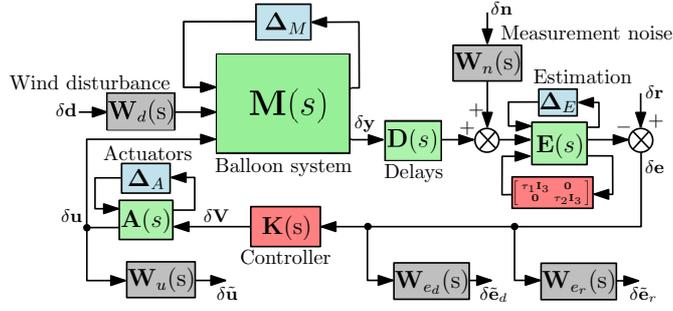}
	\caption{Closed-loop block diagram}
	 \label{fig:closed_loop}
\end{figure}

Based on the modeling presented in Sections \ref{sec:modeling} and \ref{sec:pointing}, the closed-loop system is represented in Fig.~\ref{fig:closed_loop}:
\begin{itemize}
    \item The green blocks represent the nominal system: $\mathbf M(\mathrm s)$ includes the flight chain and the pointing system, $\mathbf A(\mathrm s)=\diag \left( A_{\mathrm{el}}(\mathrm s),A_{\mathrm{ce}}(\mathrm s),A_{\mathrm{fr}}(\mathrm s) \right)$ regroups the transfer functions representing the nominal actuators dynamics, $\mathbf D (\mathrm s)$ regroups the delays, and $\mathbf E (\mathrm s)$ represents the estimation,
    \item The blue blocks $\bm \Delta_M$, $\bm \Delta_A$ and $\bm \Delta_E$ regroup the parametric uncertainties on the mechanical system, actuators' models and estimation respectively,
    \item The red blocks are the tuning parameters: controller $\mathbf K(\mathrm s) = \diag \left( K^{\mathrm{el}}(\mathrm s), K^{\mathrm{ce}}(\mathrm s), K^{\mathrm{fr}}(\mathrm s) \right)$ composed of one controller for each axis (elevation, cross-elevation, field rotation) -- the three controllers are decoupled because the action of the actuators on the line-of-sight angles are decoupled in equation \eqref{eq:angles_los} ; and the estimation parameters $\tau_1$ and $\tau_2$,
    \item The grey blocks are the weighting filters: $\mathbf W_d$ and $\mathbf W_n$ model the frequency content of the disturbances and of the noise respectively ; $\mathbf W_u$, $\mathbf W_{e_d}$ and $\mathbf W_{e_r}$ are the output weights used for the robust control synthesis,
    \item The inputs are: the wind disturbances $\delta \mathbf d \in \mathbb R^{2}$ applied along $\mathbf x$ and $\mathbf y$ to the balloon ; the sensors noises $\delta \mathbf n  \in \mathbb R^{6}$ ; and the reference angles $\delta \mathbf r  \in \mathbb R^{3}$,
    \item The internal signals are: the measured states $\delta \mathbf y  \in \mathbb R^{9}$ regrouping the 3 gondola's rotation rates, the 3 actuators angles, and the 3 angles of the line-of-sight ; the pointing error $\delta \mathbf e  \in \mathbb R^{3}$ ; the actuators voltages $\delta \mathbf V  \in \mathbb R^{3}$ ; and the commands $\delta \mathbf u = \delta \bm \theta^a \in \mathbb R^{3}$,
    \item The outputs are: the weighted errors $\delta \tilde{\mathbf e}_d \in \mathbb R^{3}$ and $\delta \tilde{\mathbf e}_r \in \mathbb R^{3}$, and the weighted command $\delta \tilde{\mathbf u} \in \mathbb R^{3}$.
\end{itemize}

The wind disturbance is characterized with the filter $D(\mathrm s)$ which was obtained in Section \ref{sec:validation}:
\begin{equation}
    \mathbf W_d(\mathrm s) = D(\mathrm s) \mathbf I_2
\end{equation}
and the measurement noise is modeled with $\mathbf W_n^{\mathrm{gyro}}$ and $\mathbf W_n^{\mathrm{optic}}$ defined in equation \eqref{eq:noise}:
\begin{equation}
    \mathbf W_n (\mathrm s) = \diag \left(  \mathbf W_n^{\mathrm{gyro}}, \mathbf W_n^{\mathrm{optic}} \right) \;.
\end{equation}

The first objective of the pointing control is disturbance rejection.
The FIREBall pointing stability must be ensured during the instrument's integration time, which is up to 100 seconds \cite{Hamden2020}.
Following the methodology presented in \cite{Ott2013}, the pointing performance is defined as the relative pointing error (RPE) across a time window $t=$ \SI{100}{\second}, accordingly to ESA standards \cite{ECSS2011a}, and picking a maximum amplitude $\epsilon_1 = \SI{5e-7}{\radian\per\hertz\tothe{0.5}}$ (elevation/cross-elevation), $\epsilon_2 = \SI{3e-5}{\radian\per\hertz\tothe{0.5}}$ (field rotation), the filter $\mathbf W_{e_d}$ is defined:
\begin{equation}
         \mathbf W_{e_d} =  \frac{t^2 \mathrm s^2 + t \sqrt{12} \mathrm s}{t^2 \mathrm s^2 + 6 t \mathrm s + 12} \diag(1/\epsilon_1, 1/\epsilon_1, 1/\epsilon_2) \;.
\end{equation}

The reference tracking requirement -- ensure a 20 rad/s (elevation and cross-elevation) or 5 rad/s (field rotation) bandwidth to follow a reference line-of-sight -- yields the filter $\mathbf W_{e_r}$ which penalizes low-frequency error:
\begin{equation}
    \mathbf W_{e_r} = \frac{1}{2} \diag \left( \frac{\mathrm s + 20}{\mathrm s + 0.02}, \frac{ \mathrm s + 20}{\mathrm s + 0.02}, \frac{ \mathrm s + 5}{\mathrm s + 0.005} \right)
\end{equation}
and since $\mathbf W_{e_r}^{-1}$ imposes an upper bound on the closed-loop sensitivity function (see equation \eqref{eq:hinfproblem}), the factor 1/2 ensures a modulus margin of at least 0.5 -- note that the sensors delays and parametric uncertainties are already taken into account in the model, so this modulus margin only needs to cover other neglected dynamics such as non-linearities of the actuators.
Finally, the limitation constraint of the actuators angles yields $\mathbf W_{u}$:
\begin{equation}
    \mathbf W_{u} = 2.5 \times 10^{5} \times \frac{\mathrm s^2 + 0.56 \mathrm s + 0.16}{\mathrm s^2 + 56 \mathrm s + 1600} \left( \frac{100 \mathrm s + 0.01}{100 \mathrm s + 1} \right)^2 \mathbf I_3
\end{equation}
which penalizes high-frequency solicitation of the actuators.
Moreover, the controller is imposed to be stable. Let us note $\bm\Delta = \diag(\bm\Delta_M,\bm\Delta_A,\bm\Delta_E)$.
The robust structured $\mathcal H_{\infty}$ problem
\begin{equation} \label{eq:hinfproblem}
\begin{array}{l}
    \underset{\mathbf K(\mathrm s), \tau_1, \tau_2}{\text{minimize}} \; \; \; \gamma_1  \; \; \; s.t. \; \; 
        \underset{{\bm \Delta}}{\max} \left\lbrace \mid \mid \begin{bsmallmatrix} \delta \mathbf d \\ \delta \mathbf n \end{bsmallmatrix} \rightarrow \delta \tilde{\mathbf e}_d \mid \mid_{\infty} \right\rbrace < \gamma_1 < 1 \\
        \text{subject to: } \left\lbrace \begin{array}{l}
        \underset{{\bm \Delta}}{\max} \left\lbrace \mid \mid \delta \mathbf r \rightarrow \delta \tilde{\mathbf e}_r \mid \mid_{\infty} \right \rbrace < \gamma_2 < 1 \\
        \underset{{\bm \Delta}}{\max} \left\lbrace \mid \mid \begin{bsmallmatrix} \delta \mathbf d \\ \delta \mathbf n \end{bsmallmatrix} \rightarrow \delta \tilde{\mathbf u} \mid \mid_{\infty} \right\rbrace < \gamma_3 < 1 \\
        \mathrm{s} \text{ is a pole of } \mathbf K \implies \mathrm{Re(s)} < 0 \\
    \end{array} \right.  \;.
\end{array} 
\end{equation}
is solved with non-smooth optimization (\textsc{Matlab} routine \texttt{systune}) to ensure the robust stability and robust performance of the closed-loop system.
Note that the weighted $\mathcal H_{\infty}$-norm is used rather than the $\mathcal H_{2}$-norm in the pointing performance (transfer $\begin{bsmallmatrix} \delta \mathbf d \\ \delta \mathbf n \end{bsmallmatrix} \rightarrow \delta \tilde{\mathbf e}_d$). Indeed, in addition to the colored noise model $\mathbf W_d(\mathrm s)$, the wind disturbance might sometimes excite the impulse response of the system. Therefore, by minimizing the resonances of the flexible modes, the $\mathcal H_{\infty}$-norm is more robust to these potential impulse disturbances than the $\mathcal H_{2}$-norm.

\subsection{Controller synthesis}

The synthesis is conducted with the following procedure. 

The estimation parameters are set to the values $\tau_1=$ \SI{0.1}{\second} and $\tau_2=$ \SI{0.5}{\second} that were used during FIREBall 2018 flight. An initial 2-nd order controller $\mathbf K_0 (\mathrm s)$ is proposed, which respects the bandwidth and actuators requirements $\gamma_2$ and $\gamma_3$, but not the disturbance rejection objective $\gamma_1$ (because of the resonances of the flexible pendulum modes of the flight chain). The performance indices are $\gamma_1=4.3$, $\gamma_2=0.90$, $\gamma_3=0.86$.

For a first synthesis, the controller is initialized with $K_0 (\mathrm s)$, and the estimation parameters are fixed (not tuned) to the values \SI{0.1}{\second} and \SI{0.5}{\second}. The 3-rd order structure is chosen for each controller $K^{\mathrm{el}}(\mathrm s), K^{\mathrm{ce}}(\mathrm s), K^{\mathrm{fr}}(\mathrm s)$, as it was found to yield satisfying results for a reasonable complexity. A controller $K_1 (\mathrm s)$ is obtained with the performance indices $\gamma_1=0.94$, $\gamma_2=0.99$, $\gamma_3=0.85$.

\begin{figure}[!ht]
	\centering
	\includegraphics[width=\columnwidth]{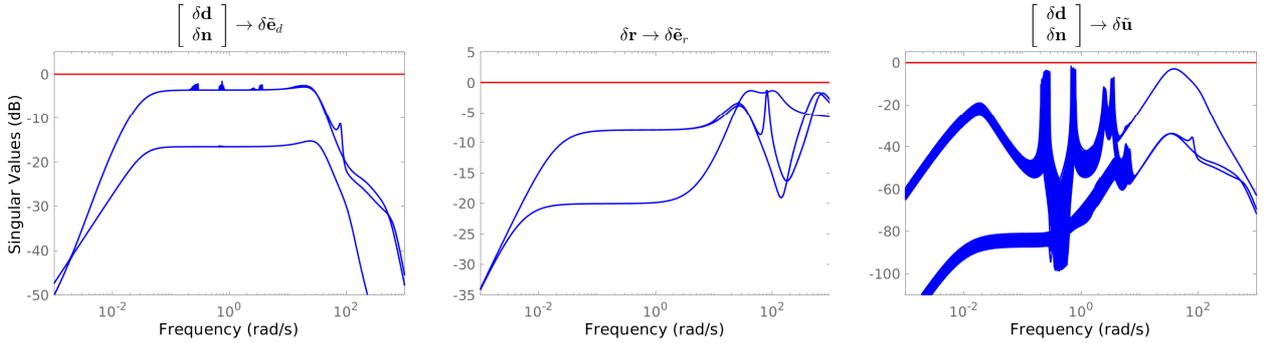}
	\caption{Singular values of the closed-loop transfers with $\mathbf K_2$ (1000 samples).}
	 \label{fig:perfo_K2}
\end{figure}

Finally, a second synthesis is initialized with $K_1 (\mathrm s)$ and the values $\tau_1=$ \SI{0.1}{\second} and $\tau_2=$ \SI{0.5}{\second}, which are now tuned simultaneously with the controller. The final values $\tau_1=$ \SI{0.066}{\second} and $\tau_2=$ \SI{94}{\second} are obtained with the controller $\mathbf K_2 (\mathrm s)$ given in equation \eqref{eq:controller}. The performance indices are $\gamma_1=0.89$, $\gamma_2=0.87$, $\gamma_3=0.85$, and the singular values of the closed-loop transfers are shown in Fig.~\ref{fig:perfo_K2}.

 \begin{equation} \label{eq:controller}
    \mathbf K_2 (\mathrm s):
    \left\lbrace \begin{array}{l}
         K_2^{\mathrm{el}}(\mathrm s) = \frac{0.36 \mathrm s^3 + 33 \mathrm s^2 + 284 \mathrm s + 4823}{\mathrm s^3 + 1054 \mathrm s^2 + 1807 \mathrm s + 665} \\
         K_2^{\mathrm{ce}}(\mathrm s) = \frac{0.95 \mathrm s^3 + 21 \mathrm s^2 + 7125 \mathrm s + 1257}{\mathrm s^3 + 999 \mathrm s^2 + 1597 \mathrm s + 0.008}  \\
         K_2^{\mathrm{fr}}(\mathrm s) =  \frac{2.78 \mathrm s^3 + 417 \mathrm s^2 + 13760 \mathrm s + 38070}{\mathrm s^3 + 1003 \mathrm s^2 + 4122 \mathrm s + 7257} \\
    \end{array} \right. 
\end{equation}

%% file: 5_conclusion.tex
\section{Conclusion}

A new general methodology is proposed to address the modeling and pointing control of an optical instrument on-board a stratospheric balloon-borne platform. 
The model of the system, based on a multibody approach, accounts for the frequency content of the wind disturbance, the dynamics of the flight chain, and the pointing system including sensors and actuators dynamics, measurement noise and line-of-sight estimation filtering. Applied to the FIREBAll mission, the nominal models fits the in-flight measurement in the frequency domain, and parametric uncertainties are included with Linear Fractional Transformations. The robust control of the line-of-sight is then tackled as a disturbance rejection problem with the $\mathcal H_{\infty}$-synthesis to minimize the relative pointing error, along with frequency-domain constraints on the sensitivity function and actuators limitation. The controller is tuned simultaneously with the line-of-sight estimation and the performance objectives are reached even in worst-case configurations of the uncertain parameters.

%% file: EGNC_paper_xyz.bbl
\begin{thebibliography}{10}

\bibitem{Yajima2009b}
N~Yajima, T~Imamura, N~Izutsu, and T~Abe.
\newblock {\em {Scientific ballooning: Technology and applications of
  exploration balloons floating in the stratosphere and the atmospheres of
  other planets}}.
\newblock Springer Science and Business Media, New-York, 2009.
\newblock
  \href{https://doi.org/10.1007/978-0-387-09727-5}{\color{blue}{\underline{DOI:\,10.1007/978-0-387-09727-5}}}.

\bibitem{Montel2019}
Johan Montel, Etienne P{\'{e}}rot, Fr{\'{e}}d{\'{e}}ri Mirc, Jean Evrard,
  Nicole Melso, and David Schiminovich.
\newblock {FIREBALL-2 (2018) in-flight pointing performance}.
\newblock In {\em 24th ESA symposium on European rocket and balloon programmes
  and related research}, pages 51--57, Essen, Germany, 2019. ESA publications.

\bibitem{Rhodes2012}
Jason Rhodes, Benjamin Dobke, Jeffrey Booth, Richard Massey, Kurt Liewer, Roger
  Smith, Adam Amara, Jack Aldrich, Joel Berge, Naidu Bezawada, Paul Brugarolas,
  Paul Clark, Cornelis~M. Dubbeldam, Richard Ellis, Carlos Frenk, Angus Gallie,
  Alan Heavens, David Henry, Eric Jullo, Thomas Kitching, James Lanzi, Simon
  Lilly, David Lunney, Satoshi Miyazaki, David Morris, Christopher Paine, John
  Peacock, Sergio Pellegrino, Roger Pittock, Peter Pool, Alexandre Refregier,
  Michael Seiffert, Ray Sharples, Alexandra Smith, David Stuchlik, Andy Taylor,
  Harry Teplitz, R.~{Ali Vanderveld}, and James Wu.
\newblock {Space-quality data from balloon-borne telescopes: The High Altitude
  Lensing Observatory (HALO)}.
\newblock {\em Astroparticle Physics}, 38(1):31--40, 2012.
\newblock
  \href{https://doi.org/10.1016/j.astropartphys.2012.05.015}{\color{blue}{\underline{DOI:\,10.1016/j.astropartphys.2012.05.015}}}.

\bibitem{Young2012}
Eliot~F. Young, Russell Mellon, Jeffrey~W. Percival, Kurt~P. Jaehnig, Jack Fox,
  Tim Lachenmeier, Brian Oglevie, and Matt Bingenheimer.
\newblock {Sub-arcsecond performance of the ST5000 star tracker on a
  balloon-borne platform}.
\newblock {\em IEEE Aerospace Conference Proceedings}, pages 1--7, 2012.
\newblock
  \href{https://doi.org/10.1109/AERO.2012.6187179}{\color{blue}{\underline{DOI:\,10.1109/AERO.2012.6187179}}}.

\bibitem{Diller2015}
Jed Diller, Kevin Dinkel, Zach Dischner, and Eliot Young.
\newblock {Design and performance of the BOPPS UVVis fine pointing system}.
\newblock {\em IEEE Aerospace Conference Proceedings}, 2015-June:1--12, 2015.
\newblock
  \href{https://doi.org/10.1109/AERO.2015.7119125}{\color{blue}{\underline{DOI:\,10.1109/AERO.2015.7119125}}}.

\bibitem{Mendillo2019}
Christopher~B. Mendillo, Kuravi Hewawasam, Glenn~A. Howe, Jason Martel,
  Timothy~A. Cook, and Supriya Chakrabarti.
\newblock {The PICTURE-C exoplanetary direct imaging balloon mission: first
  flight preparation}.
\newblock In {\em Proceedings of SPIE}, 2019.
\newblock
  \href{https://doi.org/10.1117/12.2529710}{\color{blue}{\underline{DOI:\,10.1117/12.2529710}}}.

\bibitem{Hamden2020}
Erika Hamden, D~Christopher Martin, Bruno Milliard, David Schiminovich, Shouleh
  Nikzad, Jean Evrard, Gillian Kyne, Robert Grange, Johan Montel, Etienne
  Pirot, Keri Hoadley, Donal O'sullivan, Nicole Melso, Vincent Picouet, Didier
  Vibert, Philippe Balard, Patrick Blanchard, Marty Crabill, Sandrine Pascal,
  Frederi Mirc, Nicolas Bray, April Jewell, Julia {Blue Bird}, Jose Zorilla,
  Hwei~Ru Ong, Mateusz Matuszewski, Nicole Lingner, Ramona Augustin, Michele
  Limon, Albert Gomes, Pierre Tapie, Xavier Soors, Isabelle Zenone, and Muriel
  Saccoccio.
\newblock {FIREBall-2: The Faint Intergalactic Medium Redshifted Emission
  Balloon Telescope}.
\newblock {\em The Astrophysical Journal}, 898(2), 2020.
\newblock
  \href{https://doi.org/10.3847/1538-4357/aba1e0}{\color{blue}{\underline{DOI:\,10.3847/1538-4357/aba1e0}}}.

\bibitem{Howe2017}
Glenn~A. Howe, Christopher~B. Mendillo, Kuravi Hewawasam, Supriya Chakrabarti,
  Timothy~A. Cook, Jason Martel, and Susanna~C. Finn.
\newblock {The low-order wavefront control system for the PICTURE-C mission:
  preliminary testbed results from the Shack-Hartmann sensor}.
\newblock {\em Proceedings of SPIE - The International Society for Optical
  Engineering}, (September 2017):70, 2017.
\newblock
  \href{https://doi.org/10.1117/12.2274122}{\color{blue}{\underline{DOI:\,10.1117/12.2274122}}}.

\bibitem{Romualdez2020}
L.~Javier Romualdez, Steven~J. Benton, Anthony~M. Brown, Paul Clark,
  Christopher~J. Damaren, Tim Eifler, Aurelien~A. Fraisse, Mathew~N. Galloway,
  Ajay Gill, John~W. Hartley, Bradley Holder, Eric~M. Huff, Mathilde Jauzac,
  William~C. Jones, David Lagattuta, Jason~S.Y. Leung, Lun Li, Thuy Vy~T. Luu,
  Richard~J. Massey, Jacqueline McCleary, James Mullaney, Johanna~M. Nagy,
  C.~Barth Netterfield, Susan Redmond, Jason~D. Rhodes, J{\"{u}}rgen Schmoll,
  Mohamed~M. Shaaban, Ellen Sirks, and Sut~Ieng Tam.
\newblock {Robust diffraction-limited near-infrared-to-near-ultraviolet
  wide-field imaging from stratospheric balloon-borne platforms -
  Super-pressure Balloon-borne Imaging Telescope performance}.
\newblock {\em Review of Scientific Instruments}, 91(3), 2020.
\newblock
  \href{https://doi.org/10.1063/1.5139711}{\color{blue}{\underline{DOI:\,10.1063/1.5139711}}}.

\bibitem{Fissel2013}
Laura~Marion Fissel.
\newblock {\em {PHD thesis - Probing the Role Played by Magnetic Fields in Star
  Formation with BLASTPol}}.
\newblock PhD thesis, 2013.

\bibitem{Aubin2017}
Fran{\c{c}}ois Aubin, Benjamin Bayman, Shaul Hanany, Hugo Franco, Justin Marsh,
  Joy Didier, and Amber~D. Miller.
\newblock {Torsional balloon flight line oscillations: Comparison of modelling
  to flight data}.
\newblock {\em Advances in Space Research}, 60(3):702--708, 2017.
\newblock
  \href{https://doi.org/10.1016/j.asr.2017.05.003}{\color{blue}{\underline{DOI:\,10.1016/j.asr.2017.05.003}}}.

\bibitem{Nigro1985a}
N.~J. Nigro, J.~K. Yang, A.~F. Elkouh, and N.~J. Nigro.
\newblock {Generalized math model for simulation of high-altitude balloon
  systems}.
\newblock {\em Journal of Aircraft}, 22(8):697--704, 1985.
\newblock
  \href{https://doi.org/10.2514/3.45189}{\color{blue}{\underline{DOI:\,10.2514/3.45189}}}.

\bibitem{Ward2003a}
Philip~R Ward and Keith~D DeWeese.
\newblock {Balloon borne arcsecond pointer feasibility study}.
\newblock In {\em 16th ESA Symposium on European Rocket and Balloon Programmes
  and Related Research}, volume 530, pages 197--205, St. Gallen, Switzerland,
  2003. ESA publications.

\bibitem{Quadrelli2004a}
Marco~B. Quadrelli, Jonathan~M. Cameron, and Viktor Kerzhanovich.
\newblock {Multibody dynamics of parachute and balloon flight systems for
  planetary exploration}.
\newblock {\em Journal of Guidance, Control, and Dynamics}, 27(4):564--571,
  2004.
\newblock
  \href{https://doi.org/10.2514/1.11374}{\color{blue}{\underline{DOI:\,10.2514/1.11374}}}.

\bibitem{Romualdez2018}
Luis~Javier Romualdez.
\newblock {\em {Design, Implementation, and Operational Methodologies for
  Sub-Arcsecond Attitude Determination, Control, and Stabilization of the
  Super-pressure Balloon-borne Imaging Telescope (SuperBIT)}}.
\newblock PhD thesis, 2018.

\bibitem{Kassarian2021}
E~Kassarian, F~Sanfedino, D~Alazard, H~Evain, and J~Montel.
\newblock {Modeling and stability of balloon-borne gondolas with coupled
  pendulum-torsion dynamics}.
\newblock {\em Aerospace Science and Technology}, 112:106607, 2021.
\newblock
  \href{https://doi.org/10.1016/j.ast.2021.106607}{\color{blue}{\underline{DOI:\,10.1016/j.ast.2021.106607}}}.

\bibitem{Kassarian2021a}
Ervan Kassarian, Francesco Sanfedino, Daniel Alazard, Charles-Antoine Chevrier,
  and Johan Montel.
\newblock {Linear Fractional Transformation modeling of multibody dynamics
  around parameter-dependent equilibrium}.
\newblock {\em ArXiv e-prints}, 2021.

\bibitem{Zhou1996}
Kemin Zhou, John~C Doyle, and Keith Glover.
\newblock {\em {Robust and Optimal Control}}.
\newblock Prentice hall, 1996.
\newblock
  \href{https://doi.org/10.1016/0967-0661(96)83721-X}{\color{blue}{\underline{DOI:\,10.1016/0967-0661(96)83721-X}}}.

\bibitem{DeWeese2006}
Keith~D. DeWeese and Philip~R. Ward.
\newblock {Demonstration of a balloon borne arc-second pointer design}.
\newblock {\em 36th COSPAR Scientific Assembly}, 2:10, 2006.

\bibitem{Varga2015}
Denise~M. Varga and Zach Dischner.
\newblock {Current status of a NASA high-altitude balloon-based observatory for
  planetary science}.
\newblock {\em AIAA Balloon Systems Conference 2015, MBAL 2015 - Held at the
  AIAA Aviation Forum 2015}, pages 1--9, 2015.
\newblock
  \href{https://doi.org/10.2514/6.2015-3040}{\color{blue}{\underline{DOI:\,10.2514/6.2015-3040}}}.

\bibitem{Stuchlik2015}
David~W. Stuchlik.
\newblock {The wallops arc second pointer – a balloon borne fine pointing
  system}.
\newblock {\em AIAA Balloon Systems Conference 2015, MBAL 2015 - Held at the
  AIAA Aviation Forum 2015}, (June):1--15, 2015.
\newblock
  \href{https://doi.org/10.2514/6.2015-3039}{\color{blue}{\underline{DOI:\,10.2514/6.2015-3039}}}.

\bibitem{Benford2012}
Dominic~J. Benford, Dale~J. Fixsen, Stephen~A. Rinehart, Maxime Rizzo,
  Stephen~F. Maher, and Richard~K. Barry.
\newblock {Precision attitude control for the BETTII balloon-borne
  interferometer}.
\newblock {\em Ground-based and Airborne Telescopes IV}, 8444(October
  2012):84442P, 2012.
\newblock
  \href{https://doi.org/10.1117/12.927224}{\color{blue}{\underline{DOI:\,10.1117/12.927224}}}.

\bibitem{Hawat1996}
Toufic~Michel Hawat, R.~J. Torguet, C.~Camy-Peyret, P.~Jeseck, and S.~Payan.
\newblock {Pointing and sun-tracker system for the LPMA gondola}.
\newblock {\em Proceedings of SPIE - The International Society for Optical
  Engineering}, 2739(June 1996):112--119, 1996.
\newblock
  \href{https://doi.org/10.1117/12.241908}{\color{blue}{\underline{DOI:\,10.1117/12.241908}}}.

\bibitem{Quine2002}
Brendan~M. Quine, Kimberly Strong, Aldona Wiacek, Debra Wunch, James~A. Anstey,
  and James~R. Drummond.
\newblock {Scanning the earth's limb from a high-altitude balloon: The
  development and flight of a new balloon-based pointing system}.
\newblock {\em Journal of Atmospheric and Oceanic Technology}, 19(5):618--632,
  2002.
\newblock
  \href{https://doi.org/10.1175/1520-0426(2002)019<0618:STESLF>2.0.CO;2}{\color{blue}{\underline{DOI:\,10.1175/1520-0426(2002)019<0618:STESLF>2.0.CO;2}}}.

\bibitem{Romualdez2016}
L.~J. Romualdez, P.~Clark, C.~J. Damaren, M.~N. Galloway, J.~W. Hartley, L.~Li,
  R.~J. Massey, and C.~B. Netterfield.
\newblock {Precise Pointing and Stabilization Performance for the Balloon-borne
  Imaging Testbed (BIT): 2015 Test Flight}.
\newblock {\em Proceedings of the Institution of Mechanical Engineers, Part G:
  Journal of Aerospace Engineering}, 1973, 2016.

\bibitem{Aboobaker2018}
Asad Aboobaker, Peter Ade, Derek Araujo, Fran{\c{c}}ois Aubin, Carlo
  Baccigalupi, Chaoyun Bao, Daniel Chapman, Joy Didier, Matt Dobbs, Will
  Grainger, Shaul Hanany, Kyle Helson, Seth Hillbrand, Johannes Hubmayr, Andrew
  Jaffe, Bradley Johnson, Terry Jones, Jeff Klein, Andrei Korotkov, Adrian Lee,
  Lorne Levinson, Michele Limon, Kevin MacDermid, Amber~D. Miller, Michael
  Milligan, Lorenzo Moncelsi, Enzo Pascale, Kate Raach, Britt
  Reichborn-Kjennerud, Ilan Sagiv, Carole Tucker, Gregory~S. Tucker, Benjamin
  Westbrook, Karl Young, and Kyle Zilic.
\newblock {The EBEX Balloon-borne Experiment—Gondola, Attitude Control, and
  Control Software}.
\newblock {\em The Astrophysical Journal Supplement Series}, 239(1):9, 2018.
\newblock
  \href{https://doi.org/10.3847/1538-4365/aae435}{\color{blue}{\underline{DOI:\,10.3847/1538-4365/aae435}}}.

\bibitem{Nakano2010}
Toshihiko Nakano, Yuji Sakamoto, Kazuya Yoshida, Toshinori Kuwahara, Yasuhiro
  Shoji, Makoto Taguchi, Mutumi Yamamoto, and Yukihiro Takahashi.
\newblock {The balloon-borne telescope system for optical observation of
  planets}.
\newblock {\em 2010 IEEE/SICE International Symposium on System Integration: SI
  International 2010 - The 3rd Symposium on System Integration, SII 2010,
  Proceedings}, pages 236--241, 2010.
\newblock
  \href{https://doi.org/10.1109/SII.2010.5708331}{\color{blue}{\underline{DOI:\,10.1109/SII.2010.5708331}}}.

\bibitem{Shariff2014}
J.~A. Shariff, P.~A.~R. Ade, M.~Amiri, S.~J. Benton, J.~J. Bock, J.~R. Bond,
  S.~A. Bryan, H.~C. Chiang, C.~R. Contaldi, B.~P. Crill, O.~P. Dor{\'{e}},
  M.~Farhang, J.~P. Filippini, L.~M. Fissel, A.~A. Fraisse, A.~E. Gambrel,
  N.~N. Gandilo, S.~R. Golwala, J.~E. Gudmundsson, M.~Halpern, M.~Hasselfield,
  G.~C. Hilton, W.~A. Holmes, V.~V. Hristov, K.~D. Irwin, W.~C. Jones, Z.~D.
  Kermish, C.~L. Kuo, C.~J. MacTavish, P.~V. Mason, K.~G. Megerian,
  L.~Moncelsi, T.~A. Morford, J.~M. Nagy, C.~B. Netterfield, R.~O'Brient, A.~S.
  Rahlin, C.~D. Reintsema, J.~E. Ruhl, M.~C. Runyan, J.~D. Soler, A.~Trangsrud,
  C.~E. Tucker, R.~S. Tucker, A.~D. Turner, A.~C. Weber, D.~V. Wiebe, and E.~Y.
  Young.
\newblock {Pointing control for the SPIDER balloon-borne telescope}.
\newblock {\em SPIE}, 9145(July 2014):91450U, 2014.
\newblock
  \href{https://doi.org/10.1117/12.2055166}{\color{blue}{\underline{DOI:\,10.1117/12.2055166}}}.

\bibitem{Diller2014}
Jed Diller, Kevin Dinkel, Zach Dischner, Nick Truesdale, and Eliot Young.
\newblock {Design and performance of the BBRISON UV-VIS fine pointing system}.
\newblock {\em IEEE Aerospace Conference Proceedings}, pages 1--11, 2014.
\newblock
  \href{https://doi.org/10.1109/AERO.2014.6836329}{\color{blue}{\underline{DOI:\,10.1109/AERO.2014.6836329}}}.

\bibitem{SHOJI2016}
Yasuhiro Shoji, Makoto Taguchi, Toshihiko Nakano, Atsunori Maeda, Masataka
  Imai, Yuya Gouda, Makoto Watanabe, Yukihiro Takahashi, Yuji Sakamoto, and
  Kazuya Yoshida.
\newblock {FUJIN-2:Balloon Borne Telescope for Optical Observation of Planets}.
\newblock {\em Transactions of the Japan Society for Aeronautical and Space
  Sciences, Aerospace Technology Japan}, 14, 2016.

\bibitem{Jones-Wilson2017a}
Laura Jones-Wilson, Sara Susca, Christina Diaz, Herrick Chang, Elizabeth Duffy,
  Robert Effinger, Derek Lewis, Kurt Liewer, Kevin Lo, Hared Ochoa, Joseph
  Perez, Aadil Rizvi, Carl Seubert, Carson Umsted, Michael Borden, Paul Clark,
  Richard Massey, and Michael Porter.
\newblock {A sub-arcsecond pointing stability fine stage for a high altitude
  balloon platform}.
\newblock {\em IEEE Aerospace Conference Proceedings}, 2017.
\newblock
  \href{https://doi.org/10.1109/AERO.2017.7943590}{\color{blue}{\underline{DOI:\,10.1109/AERO.2017.7943590}}}.

\bibitem{Apkarian2015}
Pierre Apkarian.
\newblock {Nonsmooth mu-synthesis}.
\newblock {\em International Journal of Robust and Nonlinear Control},
  21:1493--1508, 2010.
\newblock
  \href{https://doi.org/10.1002/rnc.1644}{\color{blue}{\underline{DOI:\,10.1002/rnc.1644}}}.

\bibitem{Apkarian2015a}
Pierre Apkarian, Minh~Ngoc Dao, and Dominikus Noll.
\newblock {Parametric Robust Structured Control Design}.
\newblock {\em IEEE Transactions on Automatic Control}, 60(7):1857--1869, 2015.
\newblock
  \href{https://doi.org/10.1109/TAC.2015.2396644}{\color{blue}{\underline{DOI:\,10.1109/TAC.2015.2396644}}}.

\bibitem{Ott2013}
T.~Ott, W.~Fichter, S.~Bennani, and S.~Winkler.
\newblock {Precision pointing H infinity control design for absolute, window-,
  and stability-time errors}.
\newblock {\em CEAS Space Journal}, 4(1-4):13--30, 2013.
\newblock
  \href{https://doi.org/10.1007/s12567-012-0028-z}{\color{blue}{\underline{DOI:\,10.1007/s12567-012-0028-z}}}.

\bibitem{ECSS2011a}
European~Space Agency.
\newblock {ESA pointing error engineering handbook, Handbook ESSB-HB-E-003},
  2011.

\bibitem{Hoblit1988}
Frederic~M. Hoblit.
\newblock {\em {Gust Loads on Aircraft: Concepts and Applications}}.
\newblock AIAA Education Series, 1988.
\newblock
  \href{https://doi.org/10.2514/4.861888}{\color{blue}{\underline{DOI:\,10.2514/4.861888}}}.

\bibitem{Defense2004}
US Department of defense.
\newblock {Flying Qualities of piloted aircraft}.
\newblock Technical report, Washington, 1990.

\bibitem{Treilhou2000}
J~P Treilhou, J~Coutelier, J~J Thocaven, and C~Jacquey.
\newblock {Payload motions detected by balloon-borne fluxgate-type
  magnetometers}.
\newblock {\em Advances in Space Research}, 26(9):1423--1426, 2000.

\bibitem{Alexander2011}
P.~Alexander and A.~{De La Torre}.
\newblock {Uncertainties in the measurement of the atmospheric velocity due to
  balloon-gondola pendulum-like motions}.
\newblock {\em Advances in Space Research}, 47(4):736--739, 2011.
\newblock
  \href{https://doi.org/10.1016/j.asr.2010.09.020}{\color{blue}{\underline{DOI:\,10.1016/j.asr.2010.09.020}}}.

\bibitem{Dubanchet2016}
Vincent Dubanchet.
\newblock {\em {Modeling and Control of a Flexible Space Robot to Capture a
  Tumbling Debris}}.
\newblock PhD thesis, Ecole Polytechnique de Montr{\'{e}}al, 2016.

\end{thebibliography}
